\documentclass[11pt]{article}
\usepackage{latexsym}
\usepackage{amssymb}
\usepackage{epsf}
\usepackage{amsmath}
\usepackage[hypertex]{hyperref}
\usepackage{graphicx}

\usepackage{slashed}

\newcommand{\tr}{{\rm Tr}}

\begin{document}
\pagestyle{empty}

\begin{flushright}
KEK-TH-1651\\
TU-941
\end{flushright}

\vspace{3cm}

\begin{center}

{\bf\LARGE Quark confinement via magnetic color-flavor locking} 
\\

\vspace*{1.5cm}
{\large 
Ryuichiro Kitano$^{1,2}$ and Naoto Yokoi$^3$
} \\
\vspace*{0.5cm}

$^1${\it 
KEK Theory Center, Tsukuba, Ibaraki 305-0801, Japan}\\
$^2${\it Department of Particle and Nuclear Physics\\
The Graduate University for Advanced Studies (Sokendai)\\
Tsukuba, Ibaraki 305-0801, Japan}\\
$^3${\it Department of Physics, Tohoku University, Sendai 980-8578,
 Japan}\\

\end{center}

\vspace*{1.0cm}

\begin{abstract}
{\normalsize
The color-flavor locking phenomenon in the magnetic picture can be the
 microscopic description of the quark confinement in QCD.
We demonstrate it in an ${\cal N} = 2$ supersymmetric $SU(N_c)_1 \times
 SU(N_c)_2$ quiver gauge theory coupled to $N_f$ flavors of quarks ($N_f
 < N_c$). 
This model reduces to $SU(N_c)_{1+2}$ gauge theory with $N_f$ flavors
 when the vacuum expectations value of the link field is much larger
 than the dynamical scales, and thus provides a continuous deformation of
 the ${\cal N}=2$ supersymmetric QCD.
We study a vacuum which survives upon adding a superpotential
 term to reduce to ${\cal N}=1$ while preserving the vectorial
 $SU(N_f)$ flavor symmetry. We find a region of the parameter space
 where the confinement is described by the Higgsing of a weakly coupled
 magnetic $SU(N_f) \times U(1)$ gauge theory.
The Higgsing locks the quantum numbers of $SU(N_f)$ magnetic color to
 those of $SU(N_f)$ flavor symmetry, and thus the massive magnetic gauge
 bosons become the singlet and adjoint representations of the flavor
 group, {\it i.e,} the vector mesons. 
If the qualitative picture remains valid in non-supersymmetric QCD, one
 can understand the Hidden Local Symmetry as the magnetic dual
 description of QCD, and the confining string is identified as the
 vortex of vector meson fields.

}
\end{abstract} 

\newpage
\baselineskip=18pt
\setcounter{page}{2}
\pagestyle{plain}
\baselineskip=18pt
\pagestyle{plain}

\setcounter{footnote}{0}

\section{Introduction}

A non-perturbative definition of asymptotically free gauge theories can,
in principle, provide us with a microscopic mechanism for the quark
confinement. The lattice QCD indeed exhibits a linear potential between
static quarks by computer simulations. The linear potential can be
interpreted as the existence of the squeezed color flux, {\it i.e.,} the
confining string.

The electric-magnetic duality in non-abelian gauge theories is another
approach to the quark confinement.
In the abelian Higgs model, a string appears as a topologically stable
configuration~\cite{Abrikosov:1956sx,Nielsen:1973cs}, which is a
magnetic flux tube. If, on the other hand, one finds the Higgs mechanism
in the magnetic picture, by duality, the string configuration
can be identified as the confining string which is a tube of a
color-electric flux~\cite{Nambu:1974zg, thooft, Mandelstam:1974vf}.
The electric-magnetic dualities in non-abelian gauge
theories~\cite{Goddard:1976qe, Montonen:1977sn} are found to exist in
various supersymmetric gauge theories~\cite{Osborn:1979tq, Sen:1994yi,
Vafa:1994tf, Seiberg:1994rs, Seiberg:1994aj, Seiberg:1994pq,
Olive:1995sw} which can be obtained as the low energy effective theories
of the branes in string/M theory~\cite{Elitzur:1997fh, Witten:1997sc}
(for a review, see \cite{Giveon:1998sr}).  Therefore, there is a hope
that the string/M theory provides a non-perturbative definition of QCD,
and the dualities equipped in the theory can explain the quark
confinement.
Indeed, it has been observed that equations to describe five-brane
configurations in M theory can be identified as the Seiberg-Witten
curves of corresponding gauge theories on the world volume, providing us
with quantum-level solutions of the gauge theories out of the classical
theories of branes~\cite{Witten:1997sc}.
The Seiberg-Witten curves know exactly where in the moduli space massless
magnetic degrees of freedom appear. The condensation of such
magnetic degrees of freedom is interpreted as the confinement.

In the real QCD, however, it seems that there is no truly weakly coupled
description of the confinement-scale physics. In order to understand the
confinement as the Higgs mechanism in the magnetic picture, we need a
deformation of the theory so that the magnetic gauge theory gets weakly
coupled and decoupled from the rest of the massive modes. Once the
deformation is done smoothly one can study physics there and then go
back to the strongly coupled regime without changing the overall
picture\footnote{See Ref.~\cite{Unsal:2007jx} for a related approach.}.

Along this line, ${\cal N}=1$ supersymmetric QCD with soft supersymmetry
breaking terms has been studied~\cite{Aharony:1995zh}. The
electric-magnetic dualities in ${\cal N}=1$ theories known as the
Seiberg duality exist for theories with $N_c + 1 < N_f < 3 N_c$,
although that is out of the region of the real QCD. It has been
observed, in the region where the Seiberg duality exists, the addition
of the soft terms triggers the spontaneous breaking of the vectorial
symmetry such as the baryon number
symmetry~\cite{ArkaniHamed:1998wc}. Since the general theorem of
Ref.~\cite{Vafa:1983tf} states that such symmetry breaking cannot occur
in non-supersymmetric QCD, it has been concluded that there is a phase
transition as the soft breaking terms get larger, and thus a simple
supersymmetric deformation of QCD did not provide a qualitative
understanding of confinement.

In another approach, Shifman and Yung have studied ${\cal N}=2$
supersymmetric $SU(N_{c})\times U(1)$ QCD with $N_{f}$ flavors ($N_{f} >
N_{c}$), by using the exact results from the Seiberg-Witten
curve~\cite{Shifman:2007kd, Shifman:2009mb, Shifman:2011ka}.  Upon
turning on the FI-term for the $U(1)$ factor, they discussed the
crossover transition from the microscopic picture to a dual magnetic
picture along the change of the FI parameter $\mu$.  In a strong
coupling vacuum for $\mu \ll \Lambda_{\rm QCD}$, a weakly-coupled dual
theory emerges as $SU(N_{f}-N_{c})\times U(1)^{N_{f}-N_{c}}$ gauge
theory with $N_{f}$ flavors.  The condensation of massless non-abelian
monopoles (or dyons) due to the FI-term causes the color-flavor locking
and produces the non-abelian vortex~\cite{Hanany:2003hp, Auzzi:2003fs}
in the dual magnetic theory, which is identified to be the confining string in
the microscopic theory~\cite{Auzzi:2003em, Shifman:2004dr,
Hanany:2004ea, Eto:2006dx}. The approach looks promising for
understanding the nature of the Seiberg duality.

Recently, a new supersymmetric deformation is proposed to possibly
connect the supersymmetric QCD and the non-supersymmetric one for $N_f <
N_c$ which is more relevant to the real world QCD~\cite{Kitano:2011zk}.
The deformation is to add $N_c$ numbers of massive flavors to the theory
so that the dual magnetic gauge group is $SU(N_f)$. The flavor symmetry
associated with the massive extra flavors, $U(N_c)$, are gauged in order
not to enhance the global symmetry of the theory.
The supersymmetry breaking terms trigger the chiral symmetry breaking
$SU(N_f)_L \times SU(N_f)_R \to SU(N_f)_V$ by the condensation of the
dual squarks while preserving the baryon number symmetry.
The same condensation breaks the dual gauge group $SU(N_f)$ completely,
and the locks the magnetic color quantum numbers to the unbroken
$SU(N_f)_V$ ones, so that the magnetic gauge boson can be identified as
the vector mesons.
The structure of the model is the same as the Hidden Local
Symmetry~\cite{Bando:1984ej} which is known to be a phenomenologically
successful model to describe the $\rho$ meson as a gauge boson of a
spontaneously broken gauge symmetry. The similarity to the real QCD can
be the sign of the smooth connection.
See also~\cite{Komargodski:2010mc, Abel:2012un} for recent works on the
interpretation of the $\rho$ meson as the gauge boson in the Seiberg
dual theory.

In the model with the extra $N_c$ flavors with the gauged $U(N_c)$
symmetry, one can observe the formation of the string associated with
the Higgsing of the dual squarks as it breaks the $U(1)$ part of the
gauged $U(N_c)$ symmetry.
However, it is not clear if this string is something to do with the
confining string as it originates from the artificially added $U(1)$
symmetry. It is also not clear what happens to the unbroken $SU(N_c)$
subgroup which is again artificially added.

In this paper, in order to clarify these issues, we study an ${\cal
N}=2$ supersymmetric extension of the model, {\it i.e.,} $SU(N_c)_1
\times U(N_c)_2$ gauge theory where $N_f$ flavors of quarks are charged
under the $SU(N_c)_1$ gauge group. There is a link field $Q$ which
transforms as the bi-fundamental representation under $SU(N_c)_1 \times
U(N_c)_2$.
Instead of adding a mass term to $Q$ as it was done in the previous
study, we consider a Higgs phase $\langle Q \rangle \neq 0$ where the
theory classically reduces to $SU(N_c)_{1+2}$ $N_f$ flavor theory.
The corresponding brane configuration can be constructed in type IIA
superstring theory, from which one can obtain the Seiberg-Witten curve by
lifting it to M theory.
One can identify the root of the Higgs branch of $Q$ in the Coulomb
branch of the moduli space, from which the theory is continuously
connected to the large $\langle Q \rangle$ region.
In the space of the root, there are points where massless monopoles
appear. Such points survive as the vacuum after a perturbation to ${\cal
N}=1$ theory, and the condensation of the monopoles by the perturbation
describes the quark confinement.
In order to further connect to non-supersymmetric QCD, one should study
points where the massless monopoles are flavor singlets since the
vector-like flavor symmetry cannot be broken in the non-supersymmetric
QCD.
We identify these points by using the curve, and study the low energy
effective theory at the points.

In a parameter region where one of the two dynamical scales is much
larger than the other, one can observe the color-flavor locking in the
magnetic picture. A non-abelian magnetic gauge group, $U(N_f)$, remains
at low energy and Higgses later by the condensation of the dual
squarks. Since the $U(1)$ factor appears as the magnetic gauge group, we
see that the color-flavor locking simultaneously describes the quark
confinement.
By the magnetic color-flavor locking, the massive magnetic gauge bosons
of $U(N_f)$ become vector mesons with flavor quantum numbers.
These are naturally identified as the $\rho$ and $\omega$ mesons in
QCD. It is interesting that in this picture a string configuration made
of $\rho$ and $\omega$ mesons can be interpreted as the confining
string~\cite{Kitano:2012zz}.

In string theory, the introduction of the extra $U(N_c)$ factor and the
link field $Q$ is somewhat a natural deformation. In the type IIA
construction, the deformation corresponds to introducing an NS5 brane
which is detached from other branes where gluons and quarks live in.
As the NS5 brane moves around, its location sets different values of $Q$
as well as the relative size of two gauge couplings (and thus the
dynamical scales), providing us with a smooth deformation of QCD.

The extra $U(N_c)$ can also be considered as the first Kaluza-Klein mode
in the sense of the dimensional deconstruction~\cite{ArkaniHamed:2001ca,
Hill:2000mu, Cheng:2001vd}.
Since the type IIA construction by D4 branes corresponds to a five
dimensional gauge theory in an interval, there are such fields by
construction. In this sense, the extra gauge factor may be a part of the
definition of QCD.

\section{A quiver model and its connection to supersymmetric QCD}

\begin{figure}[t]
\begin{center}
\includegraphics[width=7.5cm, height=5.5cm, clip]{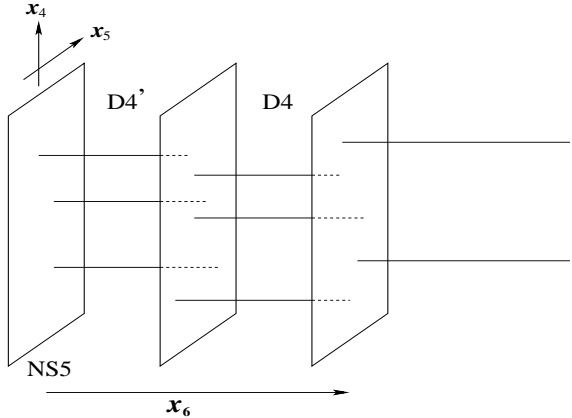}
\end{center}
\caption{Brane configuration in IIA superstring for $N_{c}=3$ and $N_{f}=2$ .}
\label{fig:brane config}
\end{figure}

\renewcommand{\arraystretch}{1.3}
\begin{table}[t]
\begin{center}
\small
 \begin{tabular}[t]{cccccc}
& $SU(N_c)_1$ & $SU(N_c)_2$ & $SU(N_f)$ & $U(1)_B$ &  $U(1)_{B^\prime}$ 
\\ \hline \hline
 $Q$ & $N_c$ & $\bar N_c$ & $1$ & 0 & 1   \\
 $F$ & $N_c$ & 1 & $N_f$ & 1 & 0 \\
\hline
 \end{tabular}
\end{center}
\caption{Quantum numbers.}
\label{tab:ele-K}
\end{table}
\renewcommand{\arraystretch}{1}

We consider an ${\cal N}=2$ supersymmetric $SU(N_{c})_{1}\times
SU(N_{c})_{2}$ gauge theory with a bi-fundamental and $N_{f}$ fundamental
hypermultiplets, whose quantum numbers are listed in
Table~\ref{tab:ele-K}.  In this paper, we assume that all the matter
fields are massless and $N_{f} < N_{c}$ for the asymptotic freedom.  The
model can be realized as a brane configuration in type IIA superstring
theory~\cite{Witten:1997sc} (see Fig.~\ref{fig:brane config}). In the
world-volume theory of the D-brane configuration, the $U(1)_{B'}$ factor
in Table~\ref{tab:ele-K} is gauged but frozen, {\it i.e.,} it is a
background field. We interpret this as taking the gauge coupling
constant of $U(1)_{B'}$ to infinity, so that the gauge multiplet is not
dynamical.

There are two limits which reduce this model to supersymmetric QCD. One
is to give a large VEV to $Q$, that corresponds to taking away the
NS5-brane in the middle of the configuration to the $x_7$, $x_8$, $x_9$
direction while connecting D4 and D4' branes. In the world-volume
theory, it corresponds to a superpotential term:
\begin{eqnarray}
 W \ni \mu^2 \Phi_{B'},
\label{eq:FI}
\end{eqnarray}
where $\Phi_{B'}$ is the chiral superfield in the (non-dynamical) gauge
multiplet of $U(1)_{B'}$. This term corresponds to the generalized
FI-term and does not break ${\cal N}=2$
supersymmetry~\cite{Hanany:1997hr, Vainshtein:2000hu}.
The $F$ and $D$-term conditions ensure the bi-fundamental $Q$ field to
obtain a VEV, and break $SU(N_{c})_{1} \times SU(N_{c})_{2} \times
U(1)_{B'}$ down to the diagonal subgroup $SU(N_{c})_{1+2}$. All of the
components in $Q$ are eaten by the gauge fields, and thus the remaining
part is supersymmetric QCD with (massless) $N_{f}$ flavors.

The other limit is to give a large VEV to $\Phi_{B'}$.  This gives a
mass to $Q$ and, at low energy, the $SU(N_{c})_{2}$ sector becomes pure
SYM and decouples from the $SU(N_{c})_{1}$ sector.  This model has been
studied in Ref.~\cite{Kitano:2011zk}.

\subsection{Seiberg-Witten curve for the quiver model}
We first obtain the solution of the quiver theory from the D-brane
configurations in string theory. By studying the Seiberg-Witten curve of
the theory, one can identify the points in the Coulomb branch of the
moduli space which are not lifted after a perturbation to an ${\cal
N}=1$ supersymmetric theory. We are particularly interested in the
vacuum where $SU(N_f) \times U(1)_B$ global symmetry is left unbroken,
since that is the case in the real world QCD.

The brane configuration in Fig.~\ref{fig:brane config} can be lifted to
an M5-brane configuration in M-theory, and the M5-brane configuration
gives the Seiberg-Witten (SW) curve~\cite{Witten:1997sc}.

For the case of $N_{f} < N_{c}-1$, the SW-curve is given by
\begin{eqnarray}
 \Lambda_2^{N_c} t^3 -
(v - \phi_1')\cdots (v - \phi_{N_c}')t^2
+
(v - \phi_1)\cdots (v - \phi_{N_c})t
- \Lambda_1^{N_c-N_f} v^{N_f}
= 0.
\end{eqnarray}
The parameters $\Lambda_{1}$ and $\Lambda_{2}$ are the dynamical scales
of $SU(N_{c})_{1}$ and $SU(N_{c})_{2}$ theories, respectively, and
$\phi_{i}$ and $\phi_{i}'$ parameterize the VEVs of the adjoint scalars
$\Phi$ and $\Phi'$ in the vector multiplets of $SU(N_{c})_{1}$ and
$SU(N_{c})_{2}$. They are subject to the traceless conditions:
\begin{eqnarray}
 \sum_{a=1}^{N_{c}} \phi'_a = \sum_{a=1}^{N_{c}} \phi_a = 0.
\label{eq: traceless}
\end{eqnarray}
For later convenience, one can represent the SW-curve in terms of gauge 
invariant Coulomb moduli $u_{i}$ and $u_{i}'$ : 
\begin{eqnarray}
 \Lambda_2^{N_c} t^3 &-&
(v^{N_c} + u'_2 v^{N_c-2} + \cdots + u'_{N_c} v^0
)t^2
\nonumber \\
&+&
(v^{N_c} + u_2 v^{N_c-2} + \cdots + u_{N_c} v^0) t
- \Lambda_1^{N_c-N_f} v^{N_f}
= 0, 
\label{eq: less than Nf-1}
\end{eqnarray}
where $u_{i}$ ($u_{i}'$) is a symmetric polynomial of $\phi_{a}$ ($\phi_{a}'$) 
with $i$-th order, which is defined by
\begin{eqnarray}
\prod_{a=1}^{N_{c}} \left( v - \phi_{a}^{(\prime)} \right) =
 \sum_{i=0}^{N_{c}} u_{i}^{(\prime)}\, v^{N_{c} - i} \ \ \ 
(u_{0} = 1),
\end{eqnarray}
where $u_1 = u_1' = 0$ by the traceless condition (\ref{eq:
traceless}).

For the special case of $N_{f} = N_{c}-1$, the SW-curve is given by
\begin{eqnarray}
 \Lambda_2^{N_c} t^3 -
(v - \phi_1')\cdots (v - \phi_{N_c}')t^2
+
(v - \hat \phi_1)\cdots (v - \hat \phi_{N_c})t
- \Lambda_1 v^{N_c - 1}
= 0, 
\end{eqnarray}
where $\hat \phi_a = \phi_a - {\Lambda_1 / N_c}$.  In terms of $\hat
\phi_a$, the trace condition is modified to\footnote{Focusing on the
$SU(N_{c})_{1}$ factor, the model corresponds to $SU(N_{c})$ QCD with
$N_{f}=2 N_{c} - 1$ flavors, where the quantum shift of the trace
condition is known to be required~\cite{Hanany:1995na, Argyres:1995wt}.}
\begin{eqnarray}
 \sum_a \phi'_a = 0,\ \ \ \sum_a \hat \phi_a = - \Lambda_1.
\label{eq: trace cond} 
\end{eqnarray}
In terms of gauge invariant moduli, the curve becomes
\begin{eqnarray}
 \Lambda_2^{N_c} t^3 &-&
(v^{N_c} + u'_2 v^{N_c-2} + \cdots + u'_{N_c} v^0
)t^2
\nonumber \\
&+&
(v^{N_c} + \Lambda_1 v^{N_c-1}
+ \hat u_2 v^{N_c-2} + \cdots + \hat u_{N_c} v^0) t
- \Lambda_1 v^{N_c - 1} = 0.
\label{eq: Nf-1}
\end{eqnarray}
Note that $\hat u_{1}$ is replaced by the dynamical scale $\Lambda_{1}$
due to the condition (\ref{eq: trace cond}).

The SW-curve (\ref{eq: less than Nf-1}) and (\ref{eq: Nf-1}) represent
the positions of M5 branes in the $(x_4, x_5, x_6, x_{10})$ space, where
$v = x_4 + i x_5$ and $t = e^{-(x_6+ix_{10})}$, which are the M-theory
lifted picture of the brane configurations in Fig.~\ref{fig:brane
config}.
The three roots of $t$ represent the positions of NS5 branes, and the
roots of $v$ represent the positions of D4 and D4'-branes.
The SW-curve encodes the low energy dynamics of our quiver gauge theory
including the full quantum mechanical effects.
   
\subsection{An Example: $SU(3)_1 \times SU(3)_2$ gauge theory with 2 flavors}
We begin with a simple and illustrative example with $N_{c}=3$ and
$N_{f}=2$, which is smoothly connected to a supersymmetric extension of
real world QCD with 2 light flavors.

The SW-curve is simply given by 
\begin{eqnarray}
\Lambda_2^3 t^3
- 
\left( v  -\phi'_1 \right) 
\left(v -\phi'_2 \right)
\left( v -\phi'_3 \right) t^2
+
\left(v - \hat \phi_1 \right)
\left(v - \hat \phi_2 \right)
\left(v - \hat \phi_3 \right) t
+
\Lambda_1 v^2
= 0,
\end{eqnarray}
where the trace condition (\ref{eq: trace cond}) applies; 
\begin{eqnarray}
 \hat \phi_1 + \hat \phi_2 + \hat \phi_3 = - \Lambda_1,\ \ \ 
 \phi'_1 + \phi'_2 + \phi'_3 = 0.
\end{eqnarray}

Using this simple example, we consider the several limits in the
SW-curve, and check that this curve completely reproduce the expected
field theory dynamics.

In the limit of $\Lambda_1 \to 0$ and $\hat \phi \to 0$, since
$SU(3)_{1}$ factor and the matter field $F$ decouple, this model should
go to $SU(3)_{2}$ gauge theory with 3 flavors, $Q$.  In this limit, the
curve becomes
\begin{eqnarray}
t \left[
\Lambda_2^3 t^2
- 
\left( v  -\phi'_1 \right) 
\left(v -\phi'_2 \right)
\left( v -\phi'_3 \right) t
+ v^3
\right]
= 0.
\end{eqnarray}
Removing $t$ factor, this curve reproduces a well-known result of the
SW-curve for $SU(3)$ theory with 3 flavors~\cite{Hanany:1995na}.  Note
that the root $t=0$ means that one of the NS5-branes goes infinitely far
away.

Next, we consider another limit: $\Lambda_2 \to 0$ and $\phi' \to 0$.
In this limit, $SU(3)_{2}$ factor decouples and our model should go to
$SU(3)_{1}$ gauge theory with 5 flavors, $Q$ and $F$.  The SW-curve is
reduced to
\begin{eqnarray}
 v^3 t^2 
- \left(v  -\hat \phi_1 \right)
\left(v  -\hat \phi_2 \right)
\left(v - \hat \phi_3 \right) t
+ \Lambda_1 v^2 = 0.
\end{eqnarray}
Simple rescaling of~ $t$~ to~ $t' = \left({v / \Lambda_1}\right)^3 t$
leads to the following factorized form,
\begin{eqnarray}
{1 \over v^3}
\left[
t'^2 
- {\left(v  -\hat \phi_1 \right)
\left(v  -\hat \phi_2 \right)
\left(v - \hat \phi_3 \right) 
\over \Lambda_1^3} t'
+ 
{ v^5 \over \Lambda_1^5 }
\right] = 0.
\end{eqnarray}
Removing $1/v^{3}$ factor, this curve reproduces the SW-curve for
$SU(3)$ QCD with 5 flavors~\cite{Hanany:1995na}.

\subsubsection{Root of the Higgs branch}
Here we consider the Higgs branch due to the VEV of the bi-fundamental
$Q$. The condensation of $Q$ leads to the diagonal breaking of quiver
gauge symmetry, $SU(3)_{1}\times SU(3)_{2} \rightarrow SU(3)_{1+2}$.
This symmetry breaking smoothly connects our quiver model to the
supersymmetric $SU(3)$ QCD with 2 flavors.

Actually, we can see this diagonal symmetry breaking in the SW-curve.
In order to handle the SW-curve, we introduce gauge invariant Coulomb
moduli $\hat u_{i}$ and $u_{i}'$:
\begin{eqnarray}
 \hat u_{2} = {\Lambda_1^2 \over 2} - {1 \over 2} (\hat \phi_1^2 + \hat \phi_2^2 + \hat \phi_3^2),\ \ \ 
 u_{2}' = - {1 \over 2} (\phi_1'^2 + \phi_2'^2 + \phi_3'^2),
\end{eqnarray}
\begin{eqnarray}
 \hat u_{3} = - \hat \phi_1 \hat \phi_2 \hat \phi_3,\ \ \ 
 u_{3}' =  - \phi_1' \phi_2' \phi_3'.
\end{eqnarray}
In terms of the above variables, the curve is
\begin{eqnarray}
 \Lambda_2^3 t^3
- \left(
v^3  
+ u_{2}' v
+ u_{3}' 
\right) t^2
+ \left[
v^3 + \Lambda_1 v^2  
+ \hat u_{2} v
+ \hat u_{3}
\right] t
- \Lambda_1 v^2 = 0.
\end{eqnarray}
When the moduli parameters satisfy the following relations,
\begin{eqnarray}
 u_{2}' = \hat u_{2},\ \ \ 
 u_{3}' - \Lambda_2^3 = \hat u_{3} ,
\label{eq:higgs}
\end{eqnarray}
the curve factorizes as 
\begin{eqnarray}
 (t - 1)\left[
\Lambda_2^3 t^2
- \left(
v^3 + \hat u_{2} v + \hat u_{3}
\right) t
+ \Lambda_1 v^2
\right] = 0.
\end{eqnarray}
The singularity at $t=1$ corresponds to a flat NS5-brane at the origin
($x_{6}=0$) in the IIA-theory language. When it is flat, taking away the
NS5-brane does not change the shape of the rest of the branes. The D4
branes in both sides of the NS5 branes are smoothly connected,
describing the Higgsing of $SU(3)_1 \times SU(3)_2 \to
SU(3)_{1+2}$~\cite{Giveon:1997sn, Erlich:1998jq}.

Removing $(t-1)$ factor gives
\begin{eqnarray}
 t'^2 - 
{\left(
v^3 + \hat u_{2} v
+ \hat u_{3} 
\right) \over \Lambda'^3 } t'
+ {v^2 \over \Lambda'^2}
=0,\ \ \ t' = (\Lambda_2 / \Lambda_1)^{3/4} t, 
\ \ \ \Lambda'^4 = \Lambda_1 \Lambda_2^3.
\label{eq:le-su3-2}
\end{eqnarray}
This is the SW-curve of the $SU(3)$ theory with 2 flavors and the
dynamical scale $\Lambda'$.
This curve gets singular when 
\begin{eqnarray}
 \hat u_{3}^2 \left(
27 \hat u_{3}^2 + 4 \hat u_{2}^3 - 24 \hat u_{2}^2 + 48 \hat u_{2} - 32
\right)
\left(
27 \hat u_{3}^2 + 4 \hat u_{2}^3 + 24 \hat u_{2}^2 + 48 \hat u_{2} + 32
\right) = 0,
\end{eqnarray}
in the unit of $\Lambda' = 1$. The curve gets maximally singular when
\begin{eqnarray}
 (\hat u_{2}, \hat u_{3}) = (\pm 2,0), \ 
\left( 
- {2 i\over \sqrt 3}, \pm {8 \sqrt 2 \over 9 \cdot 3^{1/4}} (1 + i)
\right),\ 
\left( 
 {2 i\over \sqrt 3}, \pm {8 \sqrt 2 \over 9 \cdot 3^{1/4}} (1 - i)
\right).
\label{eq:singular}
\end{eqnarray}
These are the points which survive after the mass perturbation to 
${\cal N} = 1$~\cite{Argyres:1996eh, Carlino:2000ff, Carlino:2000uk}. 
The one with $\hat u_{3} \neq 0$ provides two massless magnetic monopoles
which are singlets under flavor symmetry. See Table~\ref{tab:lowE}. 
Here $U(1)_{B'}$ factor is not listed since it is broken by taking away
the NS5 brane. The baryon number $U(1)_B$ can always be taken to be zero
by mixing with $U(1)$ gauge charges. The point is that the condensations
of the massless monopoles do not provide a massless Nambu-Goldstone
mode.

\renewcommand{\arraystretch}{1.3}
\begin{table}[t]
\begin{center}
\small
 \begin{tabular}[t]{ccccc}
& $U(1)$ & $U(1)$ & $SU(2)_f$ & $U(1)_B$  
\\ \hline \hline
 $e_1$ & $1$ & $0$ & $1$ & $0$     \\
 $e_2$ & $0$ & $1$ & $1$ & $0$     \\
\hline
 \end{tabular}
\end{center}
\caption{Massless degrees of freedom.}
\label{tab:lowE}
\end{table}
\renewcommand{\arraystretch}{1}

\subsubsection{A singular point}

We consider a special point:
\begin{eqnarray}
 \hat u_{2} = c_2 \Lambda'^2 ,\ \ \ 
\hat u_{3} = c_3 \Lambda'^3
,\ \ \ 
u_{2}' = c_2 \Lambda'^2
\ \ \ u_{3}' = \Lambda_2^3 + c_3 \Lambda'^3,
\label{eq:root}
\end{eqnarray}
where dimensionless coefficients, $(c_2, c_3)$, are one of the
combinations $(\hat u_{2},\hat u_{3})$ in Eq.~\eqref{eq:singular} except for
$(\pm 2,0)$.
This point satisfies both Eq.~\eqref{eq:higgs} and
Eq.~\eqref{eq:singular}. The low energy theory has (dual) $U(1)^2$ gauge
group and there are massless monopoles for each $U(1)$ factors as in
Table~\ref{tab:lowE}.

Upon adding masses to the adjoint chiral superfields $\Phi_1$ to reduce
the theory to ${\cal N}=1$ supersymmetry, 
\begin{align}
 W \ni& {m \over 2} \tr \Phi_1^2,
\label{eq:N=1}
\end{align}
the point in
Eq.~\eqref{eq:root} will survive as one of the vacua while most of the
point in the Coulomb branch will be lifted.
The low energy effective theory has a superpotential term:
\begin{eqnarray}
 W = - e_1 \Phi_D \bar e_1 - e_2 \Phi_D' \bar e_2 
+ m \Lambda' x \Phi_D
+ m \Lambda' x' \Phi_D'.
\end{eqnarray}
where $\Phi_D$ and $\Phi_D'$ are the singlet chiral superfields in the
gauge multiplet of $U(1) \times U(1)$. The linear terms are originated
from the perturbation to ${\cal N}=1$ where $x$ and $x'$ are some
non-zero constants. The monopoles condense through this superpotential
without breaking flavor symmetry on this vacuum, which corresponds to
one of the vacua with $r=0$, discussed in Refs.~\cite{Argyres:1996eh,
Carlino:2000ff, Carlino:2000uk}.  Therefore, the theory exhibits
confinement leaving unbroken vectorial $SU(N_f)$ and $U(1)_B$ symmetry
just as in the real world QCD.

\subsection{General $SU(N_{c})_{1}\times SU(N_{c})_{2}$ theory}

In the general $N_{c}$ and $N_{f}$ case, almost the same analysis can be
carried out.  We can easily check that the curve for $SU(N_{c})$ theory
with $N_{c}$ flavors is reproduced in the limit $\Lambda_{1}\rightarrow
0$, and the curve for $SU(N_{c})$ gauge theory with $(N_{f}+N_{c})$
flavors is reproduced in the limit $\Lambda_{2}\rightarrow 0$.

The diagonal symmetry breaking $SU(N_{c})_{1}\times SU(N_{c})_{2}
\rightarrow SU(N_{c})_{1+2}$ is realized on the following submanifold in
the moduli space.  For the case of $N_{f} < N_{c}-1$, the SW-curve
(\ref{eq: less than Nf-1}) becomes
\begin{eqnarray}
 (t-1)\left[
 \Lambda_2^{N_c} t^2 -
(v^{N_c} + u_2 v^{N_c-2} + \cdots 
+ u'_{N_c-N_f} v^{N_f} +
\cdots
+ u_{N_c} v^0) t
+ \Lambda_1^{N_c-N_f} v^{N_f}
\right]
= 0, 
\end{eqnarray}
when the following relations are satisfied:
\begin{eqnarray}
 u'_i = u_i, \ \ \ (i \neq N_c - N_f, N_c),
\end{eqnarray}
and
\begin{eqnarray}
 u'_{N_c - N_f} = u_{N_c - N_f} - \Lambda_1^{N_c - N_f},
\ \ \ 
 u'_{N_c} - \Lambda_2^{N_c} = u_{N_c}.
\end{eqnarray}
Removing $(t-1)$ factor, the above curve is reduced to
\begin{eqnarray}
 t^2 -
{v^{N_c} + u_2 v^{N_c-2} + \cdots 
+ u'_{N_c-N_f} v^{N_f} +
\cdots
+ u_{N_c} v^0
\over
\Lambda'^{N_c}}
 t
+ {v^{N_f} \over \Lambda'^{N_f}}
= 0,
\end{eqnarray}
where
\begin{eqnarray}
 \Lambda'^{2N_c-N_f} = 
\Lambda_1^{N_c-N_f}
\Lambda_2^{N_c}.
\end{eqnarray}
This is indeed the SW-curve for $SU(N_{c})$ gauge theory with $N_{f}$
flavors and the dynamical scale $\Lambda'$~\cite{Hanany:1995na,
Argyres:1995wt}.  Similarly, for $N_{f} = N_{c}-1$, the SW-curve
(\ref{eq: Nf-1}) factorizes as
\begin{eqnarray}
 (t-1)\left[
 \Lambda_2^{N_c} t^2 -
(v^{N_c} + \hat u_2 v^{N_c-2} + \cdots 
+ \hat u_{N_c} v^0) t
+ \Lambda_1 v^{N_c - 1}
\right]
= 0, 
\end{eqnarray}
for
\begin{eqnarray}
 u'_i = \hat u_i, \ \ \ (i \neq N_c),  \quad  u'_{N_c} - \Lambda_2^{N_c} = \hat u_{N_c} .
\end{eqnarray}
Removing $(t-1)$ factor leads to the SW-curve for 
$SU(N_{c})$ gauge theory with $(N_{c}-1)$ flavors~\cite{Hanany:1995na, Argyres:1995wt} 
and the dynamical scale
\begin{eqnarray}
 \Lambda'^{2N_c-N_f} = 
\Lambda_1^{N_c-N_f}
\Lambda_2^{N_c}.
\end{eqnarray}  

The singular point which is a persistent vacuum against the perturbation
to ${\cal N}=1$ also exists.  For the case of $N_{f} < N_{c} -1$,
\begin{eqnarray}
&&u'_{i} = u_{i} = c_{i} \Lambda'^{i} \ \ (i\neq N_{c}-N_{f}, N_{c}), \quad 
u'_{N_{c}} = \Lambda_{2}^{N_{c}} + u_{N_{c}} = \Lambda_{2}^{N_{c}} + c_{N_{c}} \Lambda'^{N_{c}}, \nonumber \\
&&u_{N_{c}-N_{f}} = \Lambda _{1}^{N_{c}-N_{f}} + u'_{N_{c} - N_{f}} = 
\Lambda _{1}^{N_{c}-N_{f}} + c_{N_{c}-N_{f}} \Lambda'^{N_{c}-N_{f}},
\label{eq:Root General Nf}
\end{eqnarray}
and for the case of $N_{f}=N_{c}-1$,
\begin{eqnarray}
u'_{i} = \hat u_{i} = c_{i} \Lambda'^{i} \ \  (i\neq N_{c}), \quad 
u'_{N_{c}} = \Lambda_{2}^{N_{c}} + \hat u_{N_{c}} = \Lambda_{2}^{N_{c}} + c_{N_{c}} \Lambda'^{N_{c}},
\label{eq:Root Nf=Nc-1}
\end{eqnarray} 
where $c_{i}$'s are numerical constants determined from the maximal
singularity condition.

On this special point, the low energy effective theory of the general
quiver model is a $U(1)^{N_{c}-1}$ gauge theory with the massless
monopoles charged under each $U(1)$ factor~\cite{Argyres:1996eh,
Carlino:2000ff, Carlino:2000uk}.  Quantum number of the monopoles are
listed in Table \ref{tab:General lowE}.

\renewcommand{\arraystretch}{1.3}
\begin{table}[t]
\begin{center}
\small
 \begin{tabular}[t]{ccccccc}
& $U(1)_{1}$ & $U(1)_{2}$ & $\cdots$ & $U(1)_{N_{c}-1}$ & $SU(N_{f})$ & $U(1)_B$  
\\ \hline \hline
 $e_1$ & $1$ & $0$ & $\cdots$ & $0$ & $\mathbf{1}$ & $0$     \\
 $e_2$ & $0$ & $1$ & $\cdots$ & $0$ & $\mathbf{1}$ & $0$     \\
 $\vdots$ & $\vdots$ & $\vdots$ & $\ddots$ & $\vdots$ & $\vdots$ & $\vdots$ \\
 $e_{N_{c}-1}$ & $0$ & $0$ & $\cdots$ & $1$ & $\mathbf{1}$ & $0$ \\
\hline
 \end{tabular}
\end{center}
\caption{Massless degrees of freedom in general case.}
\label{tab:General lowE}
\end{table}
\renewcommand{\arraystretch}{1}

The condensations of massless monopoles lead to complete higgsing of the
dual $U(1)^{N_{c}-1}$ gauge theory and the original theory is in the
confining phase.  Similarly to the $SU(3)$ example, all the massless
monopoles are singlet under vectorial flavor symmetry $SU(N_{f})\times
U(1)_{B}$, which is unbroken on this special point\footnote{This point
corresponds to one of the vacua with $r=0$ in
Refs.~\cite{Argyres:1996eh, Carlino:2000ff, Carlino:2000uk}}.

\section{Duality among various limits}

By using the exact results obtained from the Seiberg-Witten curve, we
now study what is happening at the points.
In this section, we study the point in Eq.~\eqref{eq:Root General Nf}
(or Eq.~\eqref{eq:Root Nf=Nc-1}) in the quantum moduli space in a field
theoretic way. The Seiberg-Witten curve told us that the low energy
theory is $U(1)^{N_{c}-1}$ gauge theory with $(N_{c}-1)$ magnetic
monopoles listed in Table~\ref{tab:General lowE} everywhere in the
parameter space of $\Lambda_1$, $\Lambda_2$ and $\mu$. Although the low
energy theory is the same everywhere, there can be variations of the
massive spectrum. We discuss below various extreme cases which allow us
to follow weakly coupled descriptions.
The field theoretic analysis makes us possible to observe the appearance
and breaking of $U(1)$ gauge factors which are important for discussion of
confinement.

\subsection{Higgs picture: $\mu \gg \Lambda_1,\  \mu \gg \Lambda_2$}
For $\mu \gg \Lambda_1$ and $\mu \gg \Lambda_2$, the classical analysis
is valid for the Higgsing of the gauge group, $SU(N_{c})_1 \times
SU(N_{c})_2 \to SU(N_{c})_{1+2}$ via the VEVs of $Q$'s:
\begin{eqnarray}
 Q = \bar Q = {i \mu \over \sqrt N_c} \cdot {\bf 1}_{N_{c}}.
\end{eqnarray}
All the $2N_c^2$ components (counting the chiral superfields) of $Q$ and
$\bar Q$ are eaten by the gauge fields of $(SU(N_{c})_1 \times SU(N_{c})_2) /
SU(N_{c})_{1+2}$ and $U(1)_{B'}$.  The low energy theory is
$SU(N_{c})_{1+2}$ theory with $N_{f}$ flavors, $F$.
The dynamical scale for the $SU(N_{c})_{1+2}$ gauge group, 
$\Lambda_{1+2}$, can be estimated by
\begin{eqnarray}
 0 = {1 \over g_1^2 (\mu)} + {1 \over g_2^2 (\mu)} 
+ {b_{1+2} \over 8 \pi^2} \log {\Lambda_{1+2} \over \mu},
\end{eqnarray}
where $\mu$ is the scale of the Higgsing. The beta function
coefficient $b_{1+2}$ is
\begin{eqnarray}
 b_{1+2} = 2N_c - N_f.
\end{eqnarray}
By using the definitions of $\Lambda_1$ and $\Lambda_2$:
\begin{eqnarray}
 0 = {1 \over g_1^2 (\mu)} 
+ {b_{1} \over 8 \pi^2} \log {\Lambda_{1} \over \mu},
\end{eqnarray}
\begin{eqnarray}
 0 = {1 \over g_2^2 (\mu)} 
+ {b_{2} \over 8 \pi^2} \log {\Lambda_{2} \over \mu},
\end{eqnarray}
with $b_1 = 2N_c - (N_f+N_c) = N_{c} - N_{f}$ and $b_2 = 2N_c - N_c =
N_{c}$, we obtain
\begin{eqnarray}
 \Lambda_{1+2} 
= (\Lambda_1^{N_c - N_f} \Lambda_2^{N_c})^{1/(2N_c - N_f)}
= \Lambda',
\end{eqnarray}
as suggested by the Seiberg-Witten curve. At the point in
Eq.~\eqref{eq:Root General Nf} or \eqref{eq:Root Nf=Nc-1}, 
low energy theory becomes a $U(1)^{N_{c}-1}$ gauge theory 
with massless monopoles which are all flavor singlets.

\subsection{Confining picture: $\Lambda_2 \gg \Lambda_1,\ \Lambda_2 \gg \mu$}
For $\Lambda_2 \gg \Lambda_1$ and $\Lambda_2 \gg \mu$, the $SU(N_c)_2$
factor gets strongly coupled before the Higgsing.  The dynamics at the
scale $\Lambda_2$ is effectively described by $SU(N_{c})_2$ theory with
$N_{c}$ flavors, $Q$.
Since $\Lambda_2 \gg \Lambda'$, the point in Eq.~\eqref{eq:Root General Nf} 
or \eqref{eq:Root Nf=Nc-1} is approximately given by
\begin{eqnarray}
u_{i}' \simeq 0 \quad (i \neq N_{c}), \ \ \ u_{N_{c}}' \simeq \Lambda_2^{N_{c}},
\end{eqnarray}
{\it i.e.,}
\begin{eqnarray}
 (\phi_1', \phi_2', \phi_3', \cdots, \phi_{N_{c}}') \simeq  
(\Lambda_2\,\omega, \Lambda_2\,\omega^2, \Lambda_2\,\omega^{3}, \cdots,
 \Lambda_2\,\omega^{N_{c}}) \times e^{i \pi/N_c}
,
\end{eqnarray}
where $\omega = \exp\left(2 \pi i/N_{c}\right)$.
This point is the baryonic root of $SU(N_{c})$  theory with $N_{c}$ 
flavors~\cite{Argyres:1996eh, Carlino:2000ff, Carlino:2000uk}.
The low energy theory is a $U(1)^{N_{c}-1}_{2}$ gauge theory with
flavor-singlet massless monopoles.\footnote{The $U(1)$ charges have some 
redundancies, as one can take different
basis of $U(1)$'s. The physics will remain unchanged in any basis.}
See Table~\ref{tab:mag-2}. 
Below the scale $\Lambda_2$, we have two
decoupled sectors: one with monopoles, $e_0$, $e_1$, $\cdots$, $e_{N_{c}-1}$, 
and the other with $SU(N_{c})_1$ gauge group with $N_{f}$ flavors of massless quarks, $F$.

Turning on the linear term \eqref{eq:FI} makes all the monopoles $e_0$,
$e_1$, $\cdots$, $e_{N_{c}-1}$ condense through the $F$- and $D$-term
conditions:
\begin{eqnarray}
 e_0 = \bar e_0 = e_1 = \bar e_1 = \cdots = \bar e_{N_{c}-1} = e_{N_{c}-1} = {\mu \over \sqrt N_c},\ \ \ 
\end{eqnarray}
The $U(1)$ gauge groups, $U(1)^{N_{c}-1}_{2} \times U(1)_{B'}$, are
all broken by the VEVs. All the monopoles are eaten by the gauge fields.

The effective dynamical scale of the $SU(N_{c})_1$ factor, $\Lambda_1'$
can be estimated as
\begin{eqnarray}
0 &=& {1 \over g_1^2 (\Lambda_0)} 
+ {b_1 \over 8 \pi^2} \log {\Lambda_2 \over \Lambda_0}
+ {b_1' \over 8 \pi^2} \log {\Lambda_1' \over \Lambda_2}
\nonumber \\
&=& {b_1 \over 8 \pi^2} \log {\Lambda_2 \over \Lambda_1}
+ {b_1' \over 8 \pi^2} \log {\Lambda_1' \over \Lambda_2},
\end{eqnarray}
where $b_1 = N_c - N_f$ and $b_1' = 2N_c - N_f$. The arbitrary
scale $\Lambda_0$ is chosen to be $\Lambda_1$ in the second line. From
this one obtains:
\begin{eqnarray}
 \Lambda_1' = (\Lambda_1^{N_c - N_f} \Lambda_2^{N_c})^{1/(2N_c - N_f)}
= \Lambda',
\end{eqnarray}
as suggested by the curve.
Since $\Lambda' \gg \Lambda_1$, the point in Eq.~\eqref{eq:Root General Nf} or \eqref{eq:Root Nf=Nc-1} 
is approximately given by
\begin{eqnarray}
 u_{i} \simeq c_{i} \Lambda'^{i} \quad \textrm{or} \quad \hat{u}_{i} \simeq c_{i} \Lambda'^{i}. 
\end{eqnarray}
This is where $(N_{c}-1)$ massless monopoles appear. The low energy theory is
again $U(1)^{N_{c}-1}$ theory with $(N_{c}-1)$ monopoles listed in Table~\ref{tab:General lowE}.

\subsection{Non-abelian magnetic picture: $\Lambda_1 \gg \Lambda_2,\
  \Lambda_1 \gg \mu$}

\renewcommand{\arraystretch}{1.3}
\begin{table}[t]
\begin{center}
\small
 \begin{tabular}[t]{ccccccccc}
& $SU(N_{c})_1$ & $U(1)^{(1)}_2$ & $U(1)^{(2)}_{2}$ & $\cdots$ & $U(1)^{(N_{c}-1)}_2$ & $SU(N_{f})$ & $U(1)_B$ 
& $U(1)_{B^\prime}$ \\ \hline \hline
 $e_0$ & ${\bf 1}$ & $1$ & $1$ & $\cdots$ & $1$ & ${\bf 1}$ & 0 & $-1$   \\
 $e_1$ & ${\bf 1}$ & $-1$ & $0$ & $\cdots$ & $0$ & ${\bf 1}$ & $0$ & $-1$   \\
 $e_2$ & ${\bf 1}$ & $0$ & $-1$ & $\cdots$ & $0$ & ${\bf 1}$ & $0$ & $-1$   \\
 $\vdots$ & $\vdots$ & $\vdots$ & $\vdots$ & $\ddots$ & $\vdots$ & $\vdots$ & $\vdots$ & $\vdots$ \\
 $e_{N_{c}-1}$ & ${\bf 1}$ & $0$ & $0$ & $\cdots$ & $-1$ & ${\bf 1}$ & $0$ & $-1$ \\
\hline
 $F$ & $N_{c}$ & $0$ & $0$ & $\cdots$ & $0$ & $N_{f}$ & $1$ & $0$   \\
\hline
 \end{tabular}
\end{center}
\caption{Effective degrees of freedom below $\Lambda_2$ for
 $\Lambda_2 \gg \Lambda_1$.}
\label{tab:mag-2}
\end{table}
\renewcommand{\arraystretch}{1}

For $\Lambda_1 \gg \Lambda_2$ and $\Lambda_1 \gg \mu$, the story is more
interesting.  At the scale $\Lambda_1$, the physics is approximately
described as the $SU(N_{c})_1$ gauge theory with $(N_{c} + N_{f})$
flavors, $Q$ and $F$.
Since $\Lambda_1 \gg \Lambda'$, the point in Eq.~\eqref{eq:Root General Nf} is
approximately, for $N_{f} < N_{c}-1$,
\begin{eqnarray}
 u_{N_{c}-N_{f}} \simeq \Lambda_1^{N_{c}-N_{f}}, \ \ \ u_{i} \simeq 0 \quad (i \neq N_{c}-N_{f}),
\end{eqnarray}
{\it i.e.,}
\begin{eqnarray}
 (\phi_1, \phi_2, \cdots, \phi_{N_{c}}) \simeq (\underbrace{0,\, 0,\, \cdots,\, 0}_{N_{f}}, 
\Lambda_1\,\omega, \Lambda_{1}\,\omega^2, \cdots,
\Lambda_{1}\,\omega^{N_{c}-N_{f}}) \times e^{i \pi / (N_c - N_f)},
\end{eqnarray}
where $\omega = \exp\left(2 \pi i/(N_{c}-N_{f})\right)$.
Similarly, for $N_f = N_c -1$,
\begin{eqnarray}
\hat u_{1} \simeq \Lambda_{1}, \ \ \ \hat u_{i} \simeq 0 \quad (i \neq 1).
\end{eqnarray}
{\it i.e.,}
\begin{eqnarray}
 (\hat \phi_1, \hat \phi_2, \cdots, \hat \phi_{N_{c}}) \simeq (\underbrace{0,\, 0,\, \cdots,\, 0}_{N_{f}}, 
-\Lambda_1).
\end{eqnarray}
These points are the baryonic root where a non-abelian magnetic gauge
group, $SU(N_{f})_1 \times U(1)^{N_{c}-N_{f}}_1$,
appears~\cite{Argyres:1996eh, Carlino:2000ff, Carlino:2000uk}.  The
effective degrees of freedom below $\Lambda_1$ is shown in
Table~\ref{tab:mag-1}.
The $SU(N_{f})_1$ factor is IR free. 
The low energy theory is again $SU(N_{c})_2$ theory with $N_{f}$ flavors, $q'$, 
but now the flavor group is replaced by the magnetic gauge group $SU(N_{f})_1$. 
The effective dynamical scale $\Lambda_2'$ for the $SU(N_{c})_2$ factor is
obtained by the equation:
\begin{eqnarray}
0 &=& {1 \over g_2^2 (\Lambda_0)} 
+ {b_2 \over 8 \pi^2} \log {\Lambda_1 \over \Lambda_0}
+ {b_2' \over 8 \pi^2} \log {\Lambda_2' \over \Lambda_1}
\nonumber \\
&=& {b_2 \over 8 \pi^2} \log {\Lambda_1 \over \Lambda_2}
+ {b_2' \over 8 \pi^2} \log {\Lambda_2' \over \Lambda_1},
\end{eqnarray}
where $b_2 = N_c$ and $b_2' = 2N_c - N_f$. From this we obtain
\begin{eqnarray}
 \Lambda_2' = (\Lambda_1^{N_c - N_f} \Lambda_2^{N_c})^{1/(2N_c - N_f)}
= \Lambda',
\end{eqnarray}
as suggested by the curve.

\renewcommand{\arraystretch}{1.3}
\begin{table}[t]
\begin{center}
\small
 \begin{tabular}[t]{cccccccccc}
& $SU(N_{f})_1$ & $U(1)^{(1)}_1$ & $U(1)^{(2)}_1$ & $\cdots$ & $U(1)^{(N_{c}-N_{f})}_1$ & $SU(N_{c})_2$ & $SU(N_{f})$ & $U(1)_B$ &  $U(1)_{B^\prime}$ 
\\ \hline \hline
 $q$ & $N_{f}$ & $1/N_{f}$ & $1/N_{f}$ & $\cdots$ & $1/N_{f}$ & ${\bf 1}$ & $N_{f}$ & $0$ & $-1$   \\
 $q'$ & $N_{f}$ & $1/N_{f}$ & $1/N_{f}$ & $\cdots$ & $1/N_{f}$ & $\bar N_{c}$ & ${\bf 1}$ & $-1$ & $0$   \\
 $e_1$ & ${\bf 1}$ & $-1$ & $0$ & $\cdots$ & $0$ & ${\bf 1}$ & ${\bf 1}$ & $0$ & $-1$ \\
 $e_2$ & ${\bf 1}$ & $0$ & $-1$ & $\cdots$ & $0$ & ${\bf 1}$ & ${\bf 1}$ & $0$ & $-1$ \\
 $\vdots$ & $\vdots$ & $\vdots$ & $\vdots$ & $\ddots$ & $\vdots$ & $\vdots$ & $\vdots$ & $\vdots$ & $\vdots$ \\
 $e_{N_{c}-N_{f}}$ & ${\bf 1}$ & $0$ & $0$ & $\cdots$ & $-1$ & ${\bf 1}$ & ${\bf 1}$ & $0$ & $-1$ \\
\hline
 \end{tabular}
\end{center}
\caption{Effective degrees of freedom below $\Lambda_1$ for $\Lambda_1
 \gg \Lambda_2$.}  \label{tab:mag-1}
\end{table}
\renewcommand{\arraystretch}{1}

Since $\Lambda' \gg \Lambda_2$, the point in Eq.~\eqref{eq:Root General
Nf} or \eqref{eq:Root Nf=Nc-1} is approximately,
\begin{eqnarray}
 u'_{i} \simeq c_{i} \Lambda'^{i} 
\end{eqnarray}
that is again where $(N_{c}-1)$ massless monopoles appear. The effective
degrees of freedom below $\Lambda'$ is listed in Table~\ref{tab:mag-3}.
Here, the $U(1)_1$ factors are rearranged (and renamed $U(1)_{\tilde
1}$) so that the charges are diagonalized. In general $U(1)_{\tilde
1}$'s are linear combinations of $U(1)_1$'s and $U(1)_2$'s. We have also
rearranged $U(1)_B$ as the one which remains unbroken at the vacuum.

\renewcommand{\arraystretch}{1.3}
\begin{table}[t]
\begin{center}
\hspace*{-0.9cm}
\small
 \begin{tabular}[t]{ccccccccccc}
& $SU(N_{f})_1$ & $U(1)^{(1)}_{\tilde{1}}$ & $\cdots$ & $U(1)^{(N_{c}-N_{f})}_{\tilde{1}}$ & $U(1)^{(1)}_2$ & $\cdots$ 
& $U(1)^{(N_{c}-1)}_2$ & $SU(N_{f})$ & $U(1)_B$ &  $U(1)_{B^\prime}$ 
\\ \hline \hline
 $q$ & $N_{f}$ & $1/N_{f}$ & $\cdots$ & $1/N_{f}$ & $0$ & $\cdots$ & $0$ & $N_{f}$ & $0$ & $-1$  \\
\hline
 $e_1$ & ${\bf 1}$ & $-1$ & $\cdots$ & $0$ & $0$ & $\cdots$ & $0$ & ${\bf 1}$ & $0$ & $-1$ \\
 $\vdots$ & $\vdots$ & $\vdots$ & $\ddots$ & $\vdots$ & $\vdots$ & $\ddots$ & $\vdots$ & $\vdots$ & $\vdots$ & $\vdots$ \\
 $e_{N_{c}-N_{f}}$ & ${\bf 1}$ & $0$ & $\cdots$ & $-1$ & $0$ & $\cdots$ & $0$ & ${\bf 1}$ & $0$ & $-1$ \\
\hline
 $e'_1$ & ${\bf 1}$ & $0$ & $\cdots$ & $0$ & $1$ & $\cdots$ & $0$ & ${\bf 1}$ & $0$ & $0$ \\
 $\vdots$ & $\vdots$ & $\vdots$ & $\ddots$ & $\vdots$ & $\vdots$ & $\ddots$ & $\vdots$ & $\vdots$ & $\vdots$ & $\vdots$ \\
 $e'_{N_{c}-1}$ & ${\bf 1}$ & $0$ & $\cdots$ & $0$ & $0$ & $\cdots$ & $1$ & ${\bf 1}$ & $0$ & $0$ \\
\hline 
 \end{tabular}
\end{center}
\caption{Effective degrees of freedom below $\Lambda'$ for $\Lambda' \gg \Lambda_2$.}  \label{tab:mag-3}
\end{table}
\renewcommand{\arraystretch}{1}

Below $\Lambda'$, the $SU(N_{f})_1$ factor gets UV free again. The
effective dynamical scale is $\Lambda'' = (\Lambda'^{N_{c}}
\Lambda_1^{N_{f}-N_{c}})^{1/N_{f}} \ll \Lambda'$.
Turning on the linear term~\eqref{eq:FI} in the region $\Lambda'' \ll
\mu \ll \Lambda'$, the $q$ and $e$ fields condense through the $F$-term
conditions:
\begin{eqnarray}
&&e'_1 \bar e'_1 = e'_2 \bar e'_2 = \cdots = e'_{N_{c}-1} \bar e'_{N_{c}-1} = 0, 
\nonumber \\ 
&&- \tr (q \bar q) - \sum_{i} e_{i} \bar e_{i} = \mu^2, \ \ \ 
q T^a \bar q  = 0, \ \ \ 
{1 \over 2} \tr (q \bar q) - \sum_{i} e_{i} \bar e_{i} = 0,
\label{eq:cfl-higgs}
\end{eqnarray}
where $T^a$ is the generator of the $SU(N_{f})_1$ group. Together with
the $D$-term conditions, the VEVs are fixed as
\begin{eqnarray}
 q = \bar q = { \mu \over \sqrt N_c} \cdot {\bf 1}_{N_{f}}, \ \ \ 
 e_{i} = \bar e_{i} = { \mu \over \sqrt N_c},\ \ \ 
 e'_{j} = \bar e'_{j} = 0 \quad (\textrm{for}~ {}^{\forall} i,\, j),
\label{eq:cfl-vev}
\end{eqnarray}
up to gauge transformations.  The gauge group $SU(N_{f})_1 \times
U(1)^{N_{c}-N_{f}}_{\tilde 1} \times U(1)_{B'}$ is broken, and $q$ and
$e$ are both eaten by the gauge fields.
The remaining gauge group and massless degrees of freedom is the same as
the ones in Table~\ref{tab:General lowE} as they should be.

The Higgsing in Eq.~\eqref{eq:cfl-vev} provides a weakly coupled
description of the magnetic color-flavor locking as well as the monopole
condensation. The unbroken $SU(N_{f})$ in Table~\ref{tab:General lowE}
is the diagonal subgroup of the magnetic gauge group $SU(N_{f})_1$ and
$SU(N_{f})$ in Table~\ref{tab:mag-3}.
In this parameter region, the massive magnetic gauge bosons of
$SU(N_{f})_1 \times U(1)^{N_{c}-N_{f}}_{\tilde 1}$ have masses of
$O(\mu)$, which is much lighter than the rest of hadrons whose masses
are around the dynamical scale $\Lambda'$.
Therefore, the quiver theory provides a continuous deformation of ${\cal
N}=2$ supersymmetric QCD, that can bring down the sector of flavored
vector mesons as a weakly coupled magnetic gauge theory.

Note that the Higgsing in Eq.~\eqref{eq:cfl-vev} leaves unbroken
$SU(N_{f})$ as well as the baryon number symmetry $U(1)_B$. Therefore, there is a
good chance that the vacuum is smoothly connected to non-supersymmetric
QCD where the spontaneous breaking of vector-like symmetries are forbidden.

\section{Deformation to ${\cal N}=1$ SUSY}

In the previous section, we have studied the vacuum which survives upon
adding a superpotential term in Eq.~\eqref{eq:N=1} to reduce to ${\cal
N}=1$ theory.
We have seen that for $\Lambda_1 \gg \Lambda_2$ and $\Lambda' > \mu$,
the effective degrees of freedom below the scale $\Lambda'$ is described
by the fields in Table~\ref{tab:mag-3}.

We here comment on the case with $\Lambda_1 \gg \Lambda_2$ and $\Lambda'
\gg m \gg \mu^2 / \Lambda'$. Since the mass term in
Eq.~\eqref{eq:N=1} is much smaller than the dynamical scale
$\Lambda'$, the analysis in the previous section remains valid. The
effective theory below the scale $\Lambda'$ has (nearly) massless
degrees of freedom as listed in Table~\ref{tab:mag-3}.
As one turns on the mass term in Eq.~\eqref{eq:N=1}, we expect that
superpotential terms are generated in the effective theory:
\begin{align}
 W &\ni 
e_i \Phi_{D \tilde 1 i} \bar e_i
- e'_i \Phi_{D2i} \bar e'_i
+ m \Lambda_1 x_{\tilde 1 i} \Phi_{D \tilde 1 i} 
+ m \Lambda' x_{2i} \Phi_{D2i},
\end{align}
at the leading order in $m$~\cite{Giveon:1997sn}.  The chiral
superfields $\Phi_{D \tilde 1 i}$ and $\Phi_{D2 i}$ are the ones in the
gauge multiplets of $U(1)_{\tilde 1}^{(i)}$ and $U(1)_2^{(i)}$,
respectively.
The dimensionless coefficients $x_{\tilde 1 i}$ and $x_{2 i}$ are
expected to be of $O(1)$.
This superpotential forces $e_i$ and $e'_i$ to condense, and breaks the
gauge symmetries, $SU(N_f)_1 \times U(1)_{\tilde 1}^{N_c-N_f} \times
U(1)_2^{N_c-1} \times U(1)_{B'}$, down to $SU(N_f)_1 \times U(1)_X$. The
superfields $e_i$ and $e'_i$ are all eaten by the gauge fields
associated with the broken generators.
The low energy effective theory below the scale $\sqrt {m \Lambda'}$ is
therefore an $SU(N_f)_1 \times U(1)_X$ gauge theory with $q$. Quantum
numbers are listed in Table~\ref{tab:q}. Here we normalized the $U(1)_X$
charge as $X = \sum_i Q_{\tilde 1}^{(i)} - B'$, where $Q_{\tilde 1}^{(i)}$ is the charge under $U(1)_{\tilde
1}^{(i)}$.

\renewcommand{\arraystretch}{1.3}
\begin{table}[t]
\begin{center}
\small
 \begin{tabular}[t]{ccccc}
& $SU(N_f)_1$ & $U(1)_X$ & $SU(N_f)$ & $U(1)_B$  
\\ \hline \hline
 $q$ & $N_f$ & $N_c/N_f$ & $N_f$ & $0$     \\
\hline
 \end{tabular}
\end{center}
\caption{Effective degrees of freedom below the scale $\sqrt {m \Lambda'}$.}
\label{tab:q}
\end{table}
\renewcommand{\arraystretch}{1}

It is important to note that the $U(1)_X$ factor is dynamical, {\it
i.e.,} the gauge coupling constant is finite. By the condensation of
$e$'s, $U(1)_{\tilde 1}^{N_c-N_f} \times U(1)_{B'}$ is broken down to a
subgroup $U(1)_X$, and the gauge coupling constant for the $U(1)_X$
gauge boson is
\begin{align}
{ 1 \over g_X^2 } & =  \sum_i {1 \over g_{\tilde 1}^{(i)2}} 
+ {1 \over g_{B'}^2}.
\end{align}
The limit $g_{B'} \to \infty$ provides finite $g_X$.

In the effective theory, there are the following terms in the
superpotential:
\begin{align}
 W & \ni - {N_c \over N_f} \tr (q \Phi_X \bar q ) + \mu^2 \Phi_X,
\end{align}
where the latter term is reduced from Eq.~\eqref{eq:FI}. The vacuum is,
therefore, at
\begin{align}
 q & = \bar q = {\mu \over \sqrt N_c}\cdot {\bf 1},
\label{eq:qvev}
\end{align}
and there the magnetic gauge group, $SU(N_f)_1$, is locked to the
$SU(N_f)$ flavor symmetry while $U(1)_X$ is spontaneously broken.
This symmetry breaking exhibits a string as a finite energy
configuration.
The string, however, should not be completely stable since the original
theory, the $SU(N_c)_1 \times SU(N_c)_2$ gauge theory, does not support
a topologically stable string. There is a $U(1)_{B'}$ factor which can
support a string, but the configuration costs infinite energy because
the gauge boson is infinitely heavy. Therefore, the string we find in
the effective theory should be broken by a pair creation of quarks which
source the magnetic flux of $U(1)_X$.
The string corresponds to the non-abelian vortex with non-abelian moduli
originated from the color-flavor locked symmetry $SU(N_{f})$. From the
matching argument of the vortex and the monopoles attached to the
endpoints~\cite{Auzzi:2003em, Eto:2006dx}, the magnetic sources in our
case should transform non-trivially under the flavor symmetry
$SU(N_{f})$.  From this viewpoint, the endpoints of our string can be
identified as original quarks, $F$ (or the composites of $F$ and $Q$).
In other words, the quarks, $F$, are confined by this string.

The lightest hadron in this regime is the massive gauge multiplets of
$SU(N_f)_1 \times U(1)_X$. Due to the color-flavor locking, the massive
gauge fields have flavor quantum numbers, adjoint and singlet, under the
$SU(N_f)$ flavor symmetry. They correspond to the vector mesons, $\rho$
and $\omega$, in QCD.
It is interesting that the Higgs mechanism for the vector mesons is the
dual picture of the quark confinement.

The quiver model we considered serves as an example of the confinement
through the magnetic color-flavor locking. Since the vacuum we studied
preserves vectorial symmetries, $SU(N_f) \times U(1)_B$, one may be able
to smoothly connect to the non-supersymmetric QCD.
As one sends the mass parameter $m$ to infinity, the model reduces to
an ${\cal N}=1$ model which has an enhanced chiral symmetry, $SU(N_f)_L
\times SU(N_f)_R$. The condensations in Eq.~\eqref{eq:qvev} then
describes the spontaneous breaking of the chiral symmetry, $SU(N_f)_L
\times SU(N_f)_R \to SU(N_f)$, which provides the massless pions.
This would be a unification of the chiral symmetry breaking and the
confinement if the qualitative picture remains the same as the case with
small $m$.
In order to connect to real QCD, we need to add masses to superparticles
and also take large $\mu$ to reduces the original gauge group $SU(N_c)
\times SU(N_c)$ to single $SU(N_c)$.
Again, although one cannot take this limit while preserving the weak
coupling of the magnetic picture, qualitative picture may remain the
same.

Indeed, it has been known that the picture of $\rho$ and $\omega$ meson
as the gauge bosons, and the pions as the uneaten Nambu-Goldstone bosons
in the Higgs mechanism for $\rho$ and $\omega$ mesons (known as the
Hidden Local Symmetry~\cite{Bando:1984ej}) is phenomenologically quite
successful. The model studied above provides a possible theoretical
reason why it is successful; the hidden gauge symmetry is actually
identified as the magnetic gauge symmetry at least in a parameter region
of a quiver and supersymmetric deformation of QCD.
As an experiment, it has been constructed a linear sigma model to
describe the Higgs mechanism for the $\rho$ and $\omega$ mesons in
Ref.~\cite{Kitano:2012zz}. The string configuration is constructed as a
solution of the classical field equations. The string tension and the
Coulomb force between monopoles are estimated by using masses and
coupling constants of hadrons as input parameters, and they are found to
be consistent with those of the QCD string.

\section{Discussion}

The quark confinement is one of the mysterious phenomena in four
dimensional gauge theories. In the real world, the confining string can
actually be seen as the linear potential between quarks which can be
inferred, for example, from the spectra of quarkonium masses. (See, e.g.,
\cite{Bali:2000gf} for a review.)
Therefore, it sounds promising that the QCD has a magnetic picture
whose Higgs mechanism supports a string as a (meta-)stable
configuration.

If there is such a picture for the confining string, there should be a
massive magnetic gauge boson in the spectrum.
Here, an interesting possibility emerges; the Higgsing of the magnetic
gauge group actually be of the color-flavor locking type so that the
lightest vector mesons, $\rho (770)$ and $\omega (782)$, are
identified as the magnetic gauge bosons.
In this picture, the vortex configuration of the $\rho$ and $\omega$
mesons is the confining string, {\it i.e.}, the flux tube of the gluon field.

We have studied the quiver model obtained from type IIA superstring
theory, that can possibly connect to the real QCD smoothly. We have
indeed observed the color-flavor locking in the magnetic picture in some
parameter regions. If the qualitative picture remains in the limit where
non-supersymmetric QCD is realized, the Hidden Local Symmetry is
understood as the magnetic picture of QCD and simultaneously describes
the confinement.

The model we studied can be thought of as the low energy theory of a
five-dimensional QCD {\it \`a la} dimensional deconstruction. For $\mu
\gg \Lambda_1$ and $\mu \gg \Lambda_2$, the Higgsing $SU(N_c)_1 \times
SU(N_c)_2 \to SU(N_c)_{1+2}$ happens in the weakly coupled regime,
corresponding to the picture of a large extra dimension where the
massive gauge boson corresponds to the Kaluza-Klein mode. As the size of
the extra dimension gets smaller, the non-perturbative effects become
important, and for $\mu \ll \Lambda_1$ and $\mu \ll \Lambda_2$ we have
seen a smooth transition to a picture of the color-flavor locking in the
magnetic picture. The extra-dimensional gauge theory provides a natural
deformation of the four-dimensional theory with the size of the
extra-dimension as the parameter to smoothly connect the weakly and
strongly coupled physics.
Such a deformation is natural in the sense of the theory of M5 brane as
the fundamental theory~\cite{ArkaniHamed:2001ie, Csaki:2002fy}.

\section*{Acknowledgements}
RK is supported in part by the Grant-in-Aid for Scientific Research
23740165 and 25105011 of JSPS. The work of NY is supported in part by
the Iwanami-Fujukai Foundation.

\end{document}